\documentclass[aps,showpacs,prb,twocolumn]{revtex4-1}

\usepackage{graphicx}
\usepackage[T1]{fontenc}
\usepackage{color}
\usepackage[dvips,colorlinks=true,bookmarks,pdftitle={},pdfauthor={},pdfsubject={},pdfkeywords={PDF,LaTeX,hyperlinks,hyperref}]{hyperref}

\newcommand{\trans}{{\cal T}}
\newcommand{\diff}{{\cal D}}
\newcommand{\ee}{\mathrm{e}}

\begin{document} 
\title{Efficient linear scaling method for computing the thermal conductivity of disordered materials}

\author{Wu Li$^{1,2}$\cite{authorship}}
\author{H\^aldun Sevin\c{c}li$^{2}$\cite{authorship}}
\author{Stephan Roche$^{2,3,4}$}
\author{Gianaurelio Cuniberti$^{2,5}$}
\email{g.cuniberti@tu-dresden.de}
\affiliation{$^1$Institute of Physics, Chinese Academy of Sciences, 100190 Beijing, China}
\affiliation{$^2$Institute for Materials Science and Max Bergmann Center of Biomaterials, Dresden
University of Technology, 01062 Dresden, Germany}
\affiliation{$^3$Institut Catal\`{a} de Nanotecnologia (ICN) and CIN2, Campus UAB, 08193 Bellaterra, Barcelona, Spain}
\affiliation{$^4$Instituci\'{o} Catalana de Recerca Avan\c{c}ats (ICREA), 08010, Barcelona, Spain}
\affiliation{$^5$Division of IT Convergence Engineering and National Center for Nanomaterials Technology, POSTECH, Pohang 790-784, Republic of Korea}

\date{\today}
\begin{abstract}
An efficient order$-N$ real-space Kubo approach is developed for the calculation of the thermal conductivity of complex disordered materials. 
The method, which is based on the Chebyshev polynomial expansion of the time evolution operator and the Lanczos tridiagonalization scheme, efficiently treats the propagation of phonon wave-packets in real-space and the phonon diffusion coefficients. 
The mean free paths and the thermal conductance can be determined from the diffusion coefficients. 
These quantities can be extracted simultaneously for all frequencies, which is another advantage in comparison with the Green's function based approaches.
Additionally, multiple scattering phenomena can be followed through the time dependence of the diffusion coefficient deep into the diffusive regime, and the onset of weak or strong phonon localization could possibly be revealed at low temperatures for thermal insulators. 
The accuracy of our computational scheme is demonstrated by comparing the calculated phonon mean free paths in isotope-disordered carbon nanotubes with Landauer simulations and analytical results. 
Then, the upscalibility of the method is illustrated by exploring the phonon mean free paths and the thermal conductance features of edge disordered graphene nanoribbons having widths of $\sim$20 nanometers and lengths as long as a micrometer, which are beyond the reach of other numerical techniques.
It is shown that,
the phonon mean free paths of armchair nanoribbons are smaller than those of zigzag nanoribbons for the frequency range which dominate the thermal conductance at low temperatures.
This computational strategy is applicable to higher dimensional systems, as well as to a wide range of materials.
\end{abstract}


\maketitle

\section{Introduction} 
Thermal conductivity of materials plays a crucial role in the efficiency of device applications at the nano-scale.
In electronic applications, it is required to transfer the excess heat effectively in order the device to work efficiently.
For thermoelectric applications, on the other hand, a low thermal conductivity is essential to achieve a high performance.
With the advances in the fabrication and manipulation at the nano-scale, it has become possible to consider new means of thermal management like tunable thermal links, \cite{thermal_link} thermal transistors \cite{transistor1,transistor2} and thermal logic gates.\cite{phononics}
Also, it is shown that the conventional barriers limiting the thermal conductivity can be broken. \cite{pernot_nature-matt}
Lying at the heart of a broad spectrum of applications with different, if not opposite, demands, thermal management at the nanoscale is attracting a growing interest.

On the other hand, carbon based materials (carbon nanotubes, graphene)~\cite{RMP,Graphene} are of special importance due to their very high electronic mobilities and thermal conductivies.\cite{balandin,RuoffScience,pop}
The extremely high thermal conductivity of two-dimensional clean graphene is not suited for energy applications.
However, a further intentional damage (with irradiation, isotope doping,...) 
or a reduction of the dimensionality (graphene nanoribbons)
of the material can offer the way to tune phonon conduction while preserving good electronic conductance. 
An ideal situation would be to design a material with exceptional electrical conduction together with a thermally insulating state. 
This turns out to be extremely challenging, and would require to find a way to strongly localize phonon modes, somehow ``trapped'' in a random disorder potential. 
The existence of an Anderson localization regime for acoustic phonons has been reported in some disordered materials~\cite{Anderson,Scheffold99}, but so far not in carbon-based materials. 
Disordered carbon nanotubes (CNT)-based bundles exhibit a tendency towards a weak thermal insulating regime~\cite{cagin,MingoCNTAL}, whereas isotope disorder remains inefficient to strongly localize coherent phonons, even in presence of a large and complex underlying disorder profile.\cite{Savic}
In contrast to carbon nanotubes, graphene nanoribbons (GNRs) offer additional sources of potential phonon scattering and localization because of the presence of irregular edges and enhanced chemical and structural reactivity that could affect low-energy phonon modes.\cite{Haldun} Other suggestions include the design of heterostructures made from CNTs\cite{Stoltz} or pristine graphene mixed with disordered (isotope impurities) GNRs~\cite{MingoPRBh}, or selective functionalization of GNRs via grafted hydrogen
impurities.\cite{baowenli}

With these advances in nano-scale fabrication and thermal management,  it becomes crucial to develop new computational techniques which are both able to account for the atomistic details of the systems and also can handle systems having sizes which are experimentally relevant.
Here, an efficient linear scaling approach to compute coherent phonon propagation in structurally complex  materials is described in detail. 
This approach is based on the Kubo methodology and on the extensive use of the MKRT technique,\cite{MKRT} which is a real-space (order-$N$)  computational framework. 
It has been successfully used for treating complex electron transport problems in quasi-periodic systems, two-dimensional disordered systems in high magnetic fields~\cite{Roche_QP}, carbon nanotubes~\cite{SRoche-Phonons,Hirose,Roche_CNT,Roche_elph,IshiiPRL}, semiconducting nanowires~\cite{Mads,LherbierNWs} or graphene-based materials.\cite{Roche_Graphene}
We note that implementation of the Lanczos method for computing the Landauer B\"uttiker conductance of low dimensional systems has also been reported.\cite{LandauerON}.

In this paper,  we first extend our recent communication on the method~\cite{Wu1} to a complete derivation of the phonon transmission coefficient starting from the original Kubo formula and within the framework of the harmonic approximation. 
This means that only elastic scattering (due for instance to isotope disorder) will be introduced, disregarding  anharmonic effects. 
The study of the dynamical properties of phonon wave-packets  will also be related to the thermal conductance which requires a phonon frequency integration over the whole spectrum.  
We then validate the method by comparison with other numerical approaches or analytical results, and further apply this method to edge disordered GNRs and discuss its limits.

\section{Computational phonon transport methodology}\label{sec:method}

The electronic transport theory in the linear response regime generally relies on the approach derived by R. Kubo.\cite{Kubo}
To investigate bulk quantum phonon transport in disordered materials, the use of the Kubo formalism turns out to be the most natural and computationally efficient one. It has already been used for investigating thermal transport in disordered binary alloys or nanocrystralline silicon.\cite{Kubophonon,Allen}
Inspired by the MKRT scheme for electron transport,\cite{MKRT} we derive a real-space implementation of the Kubo formula for phonon propagation, which establish a direct computational bridge between phonon dynamics and the thermal conductance. 
In contrast to other implementations of the Kubo approach, we extract the dynamical information from the time evolution of the wave-packet~\cite{WPP} (based on the expansion of the evolution operator on a Chebyshev polynomials basis) and simulate quantum dynamics 
instead of solving the Newtonian equations of motion.\cite{srivastava,Kubo33,loh}
Therefore, a single initial condition is enough, i.e. the initial atomic displacements, without any need to compute the time-dependent atomic velocities. 
Additionally, by using the Lanczos technique, one can avoid any matrix inversions, and a considerable gain in the computational efficiency is obtained, which allows the study of very large scale materials.

\subsection{Derivation of the phonon transport equations}
The vibrational Hamiltonian, taking only the harmonic interactions into account, is described as 
\begin{equation}
\label{halmiltonian}
\mathcal{H}=\sum_{i} \frac{\hat{p}_{i}^{2}}{2M_{i}}+\sum_{ij} \Phi_{ij}\hat{u}_{i}\hat{u}_{j},
\end{equation}
where $\hat{u}_{i}$ and $\hat{p}_{i}$ are the displacement and momentum operators for the $i$th atomic degree of freedom, $M_{i}$ is the corresponding mass, and $\Phi$ is the force constant tensor. 
Based on the linear response theory, the phonon conductivity $\sigma$ along $x$ direction can be obtained as~\cite{Allen}
\begin{equation}\label{kubothermal}
\sigma=\Omega T^{-1}\lim_{\omega\to0}\lim_{\eta\to0}\int_{0}^{\beta}d\lambda\int_{0}^{\infty}dt\,\ee^{i(\omega+i\eta)t}\langle \hat{J}^{x}(-i\hbar\lambda)\hat{J}^{x}(t)\rangle,
\end{equation}
with $\Omega$ being the system volume and $T$ being the temperature. $\hat{J}^{x}$ is the $x$ component of the energy flux operator $\hat{\textbf{J}}$, and it can be expressed as $\hat{J}^{x}={1}/{2\Omega}\sum_{ij} (X_{i}-X_{j})\Phi_{ij}\hat{u}_{i}\hat{v}_{j}$, where $\hat{v}_{j}$ is the velocity operator and $X_{i}$ is the equilibrium position of the atom to which the $i$th degree of freedom belongs. 
After neglecting terms like $\hat{a}_m^{\dagger}\hat{a}_n^\dagger$ and $\hat{a}_m\hat{a}_n$, $\hat{J}^{x}$ can be rewritten in terms of the phonon creation and annihilation operators as $\hat{J}^{x}=\sum_{m,n}J^x_{mn}\hat{a}_m^{\dagger}\hat{a}_n$, where 
\begin{equation} \label{HeatOperator}
 J^x_{mn}=
-\frac{i\hbar}{4\Omega}
\left(
\sqrt{\frac{\omega_m}{\omega_n}}+
\sqrt{\frac{\omega_n}{\omega_m}}
\right)
\langle m|[X,D]|n \rangle.
\end{equation}
$X$ is the diagonal matrix of equilibrium positions, $D$ is the mass normalized dynamical matrix with  $D_{ij}=\Phi_{ij}/{\sqrt{M_{i}M_{j}}}$ 
and $|n\rangle$ is the $n$th eigenstate of $D$. 
Allen and Feldman have shown that Eq.(\ref{kubothermal}) can be written as \cite{Allen} 
\begin{equation}
  \sigma=\frac{\pi \Omega}{\hbar T}\sum_{m,n}\frac{\partial f_B}{\partial\omega_m}J^x_{mn}J^x_{nm}\delta(\omega_m-\omega_n).
\end{equation}
where $f_B$ is the Bose distribution function. 
Therefore, one has
\begin{equation}
\sigma=-\frac{\pi}{\Omega}\int\limits_{0}^{\infty}
\mathrm{d}\omega\,\frac{\hbar}{4\omega}
\frac{\partial f_B}{\partial T}
\textrm{Tr}\{[\hat{X},D]\delta(\omega-\sqrt{D})[\hat{X},D]\delta(\omega-\sqrt{D})\},
\end{equation}
where $\sqrt{D}=\sum_n\omega_n|n\rangle\langle n|$ acts as the free particle Hamiltonian, and $\mathrm{Tr\{\dots\}}$ stands for the trace operator.
Defining ${V}_x=-i[X,\sqrt{D}]$, one can write the thermal conductance of a one-dimensional system as
\begin{equation}
\kappa=\frac{\pi}{L^{2}}
\int_{0}^{\infty}\mathrm{d}\omega\,\hbar\omega
\frac{\partial f_B}{\partial T}
\textrm{Tr}\{V_{x}\delta(\omega-\sqrt{D})V_{x}\delta(\omega-\sqrt{D})\}.
\end{equation}
The thermal conductance can also be derived from the Landauer formalism \cite{Rego_Mingo} or  the nonequilibrium Green's function approach \cite{ciraci_yamamoto} as
\begin{equation}
\kappa=
\frac{1}{2\pi}
\int_{0}^{\infty}\mathrm{d}\omega\,\hbar\omega
\frac{\partial f_B}{\partial T}
\trans(\omega),
\label{kappa}
\end{equation}
with $\trans(\omega)$ being the phonon transmission function. Comparing these two formulas, we obtain the transmission function as
\begin{equation}\label{KuboTrans}
\trans(\omega)=\frac{2\pi^2}{L^{2}}\textrm{Tr}\{V_{x}\delta(\omega-\sqrt{D})V_{x}\delta(\omega-\sqrt{D})\}.
\end{equation}
The phonon transmission function derived here has exactly the same form as the electron transmission function derived from the Kubo-Greenwood formula~\cite{MKRT},
\begin{equation}\label{KuboTransEl}
\trans_{\mathrm{el}}(E)=
\frac{2\pi^2\hbar^{2}}{L^{2}}\textrm{Tr}\{\hat{V}_{x}\delta(E-{\hat{\mathcal{H}}}_\mathrm{el})\hat{V}_{x}\delta(E-{\hat{\mathcal{H}}}_\mathrm{el})\},
\end{equation} 
where $\hat{\mathcal{H}}_\mathrm{el}$ is the electronic Hamiltonian.

We rewrite the Kubo formula in a more convenient form for the real-space study of the wave-packet dynamics.
Expressing the $\delta-$function as 
\begin{equation}
\delta(\omega-\sqrt{D})=\frac{1}{2\pi}\int_{-\infty}^{\infty}\mathrm{d}t\,\ee^{i(\omega-\sqrt{D})t},
\end{equation}
one can then rewrite  $\trans(\omega){L^2}/{\pi}$  as
\begin{eqnarray}\label{vx0vxt}
    &&\int_{-\infty}^{\infty}\mathrm{d}t\mathrm{Tr}\left\{\delta(\omega-\sqrt{D})\right\}\langle V_x(t)V_x(0)\rangle_{\omega}\nonumber\\
    &=&\int_{0}^{\infty}\mathrm{d}t\mathrm{Tr}\left\{\delta(\omega-\sqrt{D})\right\}\langle V_x(t)V_x(0)+V_x(0)V_x(t)\rangle_{\omega}
\end{eqnarray}
where $V_x(t)=U^\dag(t)V_xU(t)$ with $U(t)=\ee^{-i\sqrt{D}t}$, and $\langle\dots\rangle_{\omega}$ denotes the mean value of the operator over different eigenstates with frequency $\omega$.
The mean-square displacement along the $x-$direction over the states having frequency $\omega$ is written as
\begin{equation}\label{chi}
	\chi^2(\omega,t)=\langle (X(t)-X(0))^2\rangle_{\omega},
\end{equation}
where $\chi(t)=U^\dag(t)\chi U(t)$. Hence, 
\begin{eqnarray}
  \frac{\mathrm{d}}{\mathrm{d}t}\chi^2(\omega^2,t)&=&
  \langle V_x(0)(X(0)-X(-t))\nonumber\\
  &&+(X(0)-X(-t))V_x(0)\rangle_{\omega},
\end{eqnarray}
and
\begin{eqnarray}
    \frac{\mathrm{d}^2}{\mathrm{d}t^2}\chi^2(\omega^2,t)=
    \langle V_x(t)V_x(0)+V_x(0)V_x(t)\rangle_{\omega}.
\end{eqnarray}
Using $\frac{\mathrm{d}}{\mathrm{d}t}\chi^2(\omega^2,t)|_{t=0}=0$, the integral in (\ref{vx0vxt}) can be evaluated as
\begin{equation}
\trans(\omega)=\frac{2\omega\pi}{L^2}\mathrm{Tr}\left\{\delta(\omega^2-D)\right\}\lim_{t\to\infty}\frac{\mathrm{d}}{\mathrm{d}t}\chi^2(\omega,t).
\end{equation}
We consider 
$\langle V_x(0)V_x(t) \rangle_\omega =v^2(\omega)\ee^{-t/\tau_{\mathrm{tr}}}$, where $v(\omega)$ and $\tau_\mathrm{tr}$ stand for the average group velocity and the average transport time at frequency $\omega$, respectively.
With this approximation, the multiple scattering effects are taken into account in an average manner, and the mean-square displacement is obtained as
\begin{equation}
	\chi^2(\omega,t)=
	2v^2\tau_\mathrm{tr}t 
	- 2v^2\tau_\mathrm{tr}^2 
	+ 2v^2\tau_\mathrm{tr}^2 \ee^{-t/\tau_\mathrm{tr}}.
\end{equation}
The diffusion coefficient is defined as
\begin{equation}\label{diffusion}
    \mathcal{D}(\omega,t)=\frac{\chi^2(\omega,t)}{t},
\end{equation}
therefore in the ballistic regime $t\ll\tau_\mathrm{tr}$ and $\mathcal{D}(\omega,t)=v^2t$, while in the diffusive regime $t\gg\tau_\mathrm{tr}$ and  $\mathcal{D}(\omega,t)=\mathcal{D}_\mathrm{max}(\omega)=2v^2\tau_\mathrm{tr}$.
The transmission function in the diffusive regime reduces to
\begin{equation}\label{Trans-diffusive}
  \trans(\omega)=\frac{2\omega\pi}{L^2}\mathrm{Tr}\left\{\delta(\omega^2-D)\right\}\mathcal{D}_{max}(\omega).
\end{equation}
The phonon transport mean free path (MFP) can be obtained via
\begin{equation}\label{mfp}
  \ell(\omega)=v\tau_\mathrm{tr}=\frac{\mathcal{D}_\mathrm{max}(\omega)}{2v(\omega)}.
\end{equation}
We note that,
$\ell(\omega)$ can also be approximated by $\trans(\omega)L/2N_\mathrm{ch}$, which gives the same result as Eq.(\ref{mfp}), because in a quasi-one dimensional system the number of channels is $N_\mathrm{ch}(\omega)\approx2\omega\pi\mathrm{Tr}\left\{\delta(\omega^2-D)\right\}v(\omega)/L$.

By computing $\mathcal{D}(\omega,t)$, one can thus deduce $\mathcal{D}_\mathrm{max}(\omega)$ and $v(\omega)$ and $\ell(\omega)$. 
Noticing that 
\begin{eqnarray}\label{chiformal}
 \chi^2(\omega,t)&=&\frac{\mathrm{Tr}\left\{(X(t)-X(0))^2\delta(\omega^2-D)\right\}}{\mathrm{Tr}\left\{\delta(\omega^2-D)\right\}}\nonumber\\
                 &=&\frac{\mathrm{Tr}\{[X,U(t)]^{\dag}\delta(\omega^2-D)[X,U(t)]\}}{\mathrm{Tr}\left\{\delta(\omega^2-D)\right\}},
\end{eqnarray}
the trace in the numerator can be calculated efficiently through an average over a few random phase states as
$N\overline{\langle\psi|[X,U(t)]^{\dag}\delta(\omega^2-D)[X,U(t)]|\psi\rangle}$,
$N$ being the number of degrees of freedom and the bra-ket corresponds to a kind of projected density of states (${\widetilde{\rm PDOS}}$) associated with the vector $[X,U(t)]|\psi\rangle$. 
${\widetilde{\rm PDOS}}$ differs from the normal PDOS by a factor of $1/2\omega$ and is obtained by using the Lanczos method.
$\mathrm{Tr}\left\{\delta(\omega^2-D)\right\}$ is also calculated by averaging the
 ${\widetilde{\rm PDOS}}$ of $|\psi\rangle$ through the Lanczos method.
$[X,U(t)]|\psi\rangle$ is calculated by using Chebyshev expansion of $U(t)$, however the very low frequency component cannot be evaluated with a reasonable accuracy for large $t$, which will be discussed in Section \ref{subsec:chebyshev}.
If we transform $U(t)\to\mathcal{U}(\tau)=\mathrm{e}^{-iD\tau}$, the corresponding velocity and the transport time will be transformed as $v\to2\omega v$ and $\tau_\mathrm{tr}\to\tau_\mathrm{tr}/2\omega$, up to the first order approximation of the perturbation of the dynamical matrix $D$. Therefore Eq.(\ref{chiformal}) can be approximated by 
\begin{equation}
 \chi^2(\omega,t)\approx\frac{\mathrm{Tr}\{[X,\mathcal{U}(t/2\omega)]^{\dag}\delta(\omega^2-D)[X,\mathcal{U}(t/2\omega)]\}}{\mathrm{Tr}\left\{\delta(\omega^2-D)\right\}},
\end{equation}
so that we can expand $\mathcal{U}(\tau)$ instead of $U(t)$ in terms of the Chebyshev polynomials.

\subsection{Lanczos and continued fraction methods}
Efficient computational recursion and order-$N$ methods have been successfully developed in solid-state physics starting from the pioneering work by R. Haydock.\cite{Recursion1,Recursion2,ON} The recursion methods are based on an eigenvalue approach of Lanczos~\cite{Lanczos}, and rely on the computation of Green's function matrix elements by continued fraction expansion, which can be implemented either in real or reciprocal spaces. These techniques are particularly well suited for treating disordered materials (alloys,..) and defect-related problems, and were successfully implemented to tackle impurity-level calculations in semiconductors using the tight-binding approximation~\cite{Recsi}, or electronic structure investigations of amorphous semiconductors, transition metals and metallic glasses based on the linear-muffin-tin orbitals.\cite{RecLMTO} The vibrational DOS of disordered materials have also been investigated by the recursion method.\cite{DoSv}

\begin{figure}[t]
\begin{center}\leavevmode
\includegraphics[width=84mm]{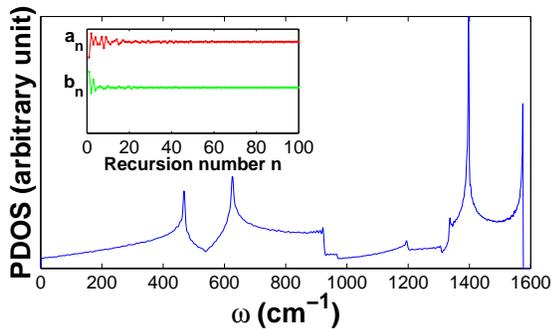}
\caption{(Color online) Vibrational ${{\rm PDOS}}$ on a single random phase state for a clean two-dimensional graphene system calculated by using the force constants given in Ref.~\onlinecite{zimmermann}.The Lanczos coefficients are shown in the inset.}
\label{Dos_Graphene}
\end{center}
\end{figure}

Owing to the $\delta-$function, the numerator and the denominator can be considered as ${\widetilde{\rm PDOS}}$ on $|\psi\rangle$ and $[X,U(t)]|\psi\rangle$, respectively. Given $|\psi\rangle$ and assuming that $[X,U(t)]|\psi\rangle$ is already known, we can use continued fraction method to calculate ${\widetilde{\rm PDOS}}$. We tridiagonalize the dynamical matrix $D$ from an initial state
 $|\psi\rangle$ or $[X,U(t)]|\psi\rangle$ by using the Lanczos scheme. The Lanczos coefficients $a_n$ and $b_n$ display a convenient feature: they converge rapidly to constants $a_{\infty}$ and $b_{\infty}$, respectively for a reasonable number of recursion steps.
 The ${\widetilde{\rm PDOS}}$ is then deduced from the first diagonal element of retarded Green's function in the 
tridiagonalized representation. For instance, 
\begin{equation}
  \langle\psi|\delta(\omega^2-D)|\psi\rangle=-\frac{1}{\pi}\lim_{\eta\to0}\mathrm{Im}\langle\psi|\frac{1}{\omega^2+i\eta-D}|\psi\rangle,
\end{equation}
and $\lim_{\eta\to0}\langle\psi|(\omega^2+i\eta-D)^{-1}|\psi\rangle$ can be expressed as a continued fraction
\begin{equation}
\frac{1}{\omega^2+i\eta-a_1-\frac{\displaystyle b_1^2}{\displaystyle \omega^2+i\eta-a_2-\frac{b_2^2}{\frac{\ddots}{\displaystyle \omega^2+i\eta-a_N-b_N^2\Sigma(\omega)}}}}
\end{equation}
In practice, the continued fraction is terminated after a certain step $N$ by the approximation $a_{N+1}=a_{\infty}$ and $b_{N+1}=b_{\infty}$ . The termination can be analytically obtained as
\begin{equation}
\Sigma(\omega)=\frac{\omega^2+i\eta-a_{\infty}-i\sqrt{(2b_{\infty})^2-(\omega^2+i\eta-a_{\infty})^2}}{2b_{\infty}^2}
\end{equation}

As an example, the normal PDOS on a single random phase state for a clean two-dimensional graphene system is given 
in Fig.~\ref{Dos_Graphene}. It already agrees well with the DOS obtained by direct matrix diagonalization. 
The inset shows the corresponding Lanczos coefficients, $a(n)$ and $b(n)$.

\subsection{Chebyshev polynomial expansion}\label{subsec:chebyshev}
The evolution of the vector $[X,U(t)]|\psi\rangle$ is determined by $U(t)$. 
Although $U(t)$ has a simple form, its exact calculation requires the diagonalization of the dynamical matrix.
Provided that all the eigenvalues and eigenvectors of $D$ are known, the matrix from of $U(t)$ can be obtained by using a unitary transformation.
The diagonalization of a matrix requires $\textit{O}(N^{3})$ operations so it is practically impossible to handle a system with size $N\simeq 10^6$.
One alternative is the Taylor expansion of the evolution operator.
The efficiency and the accuracy are however difficult to fulfill at the same time with such an  expansion, especially because a higher accuracy requires either a larger number of time steps or larger  expansion orders for a given time $t$. 
The implementation of the Taylor expansion for studying long time propagation of wave-packets in large scale disordered systems 
with a high accuracy is therefore beyond today's computational reach. 
Instead, we employ the Chebyshev polynomial expansion approach. 
Formally we can expand any function of an operator  in terms of Chebyshev polynomial-based operators, since the set of Chebyshev polynomials $Q_n$ form a complete orthogonal basis set. 
The expansion coefficients depend on the form of the expanded function, on the time step $\Delta t$, and on the function domain $[a-2b,a+2b]$ which should cover the whole spectrum of the dynamical matrix $D$. 
Since Chebyshev polynomials are defined in the interval $[-1,1]$, we first need to scale and shift the argument so that it falls
within the range based on $D'=(D-a)/2b$. Finally $U(\Delta t)$ can be expanded as 
\begin{equation}\label{chebyshev series}
	U(\Delta t)=\ee^{-i\Delta t\sqrt{a+2bD'}}=\sum_{n=0}^{\infty}c_{n}(\Delta t)Q_{n}(D')
\end{equation}
where 
\begin{equation} \label{chebyshev coefficents}
c_{n}(\Delta t)=\frac{2}{\pi(1+\delta_{n,0})}\int_{-1}^{1}dx'\frac{1}{\sqrt{1-x'^2}}\ee^{-i\Delta
t\sqrt{a+2bx'}}Q_{n}(x')
\end{equation}
Thus $[X,U(\Delta t)]|\psi \rangle = \sum_{n=0}^{\infty} c_{n}(\Delta t) [X,Q_{n}(D
)]|\psi \rangle $. Then, the commutators are computed using the Chebyshev
recurrence relation  $bQ_{n+1}(D')=(D-a)Q_{n}(D')-bQ_{n-1}(D' )$ where $Q_{0}(D')=1$
and $Q_{1}(D')=(D-a)/2b$. Recurrence relations between commutators write
$b[X,Q_{n+1}(D')]=[X,(D-a)Q_{n}(D')]-b[X,Q_{n-1}(D')]$, which is rewritten for
convenience as 
\begin{eqnarray} 
  |\alpha_{n}\rangle = Q_{n}(D')|\psi \rangle , \qquad
  |\beta_{n}\rangle = [X,Q_{n}(D')]|\psi \rangle .
\end{eqnarray}

Using the well-known expression  $[A,BC] = [A,B]C + B[A,C]$, the commutator
relationship becomes $b|\beta_{n+1}\rangle = (D-a)|\beta_{n}\rangle -
b|\beta_{n-1}\rangle + [X,D ]|\alpha_{n}\rangle$ with $|\beta_{0}\rangle = 0$ and 
$|\beta_{1}\rangle = [X,D ]|\psi \rangle /2b$.
The computation of $|\beta_{n}\rangle$ require the knowledge of vectors $|\alpha_{n}\rangle
= Q_{n}(D')|\psi \rangle$ that will appear in the Chebyshev expansion of the
evolution operator  $U(\Delta T)|\psi \rangle$. Such $|\alpha_{n}\rangle$ vectors
satisfy
\begin{equation}
 b|\alpha_{n+1}\rangle = (D-a)|\alpha_{n}\rangle - b|\alpha_{n-1}\rangle ,
\label{eq:rec_alpha}
\end{equation}
with $|\alpha_{0}\rangle = |\psi \rangle$ and $|\alpha_{1}\rangle = (D -a)|\psi\rangle /2b$. 
The algorithm thus consists of computing in parallel two recurrence relations and summing up the series
\begin{eqnarray}
    U(\Delta t)|\psi \rangle = \sum_{n=0}^{N_\mathrm{poly}} c_{n}(\Delta t)|\alpha_{n}\rangle,
\end{eqnarray}
and
\begin{eqnarray}
    [X,U(\Delta t)] |\psi \rangle = \sum_{n=0}^{N_\mathrm{poly}} c_{n}(\Delta t)|\beta_{n}\rangle.
\end{eqnarray}
In order to reach the desired accuracy, the number of recursion steps ($N_\mathrm{poly}$) needs to be chosen appropriately depending on the evolution step and the spectral bandwidth.
Once $U(m\Delta t)|\psi \rangle$
and $[X,U(m\Delta t)]|\psi \rangle$ are obtained at time $m\Delta t$, then for
the following evolution step one has
\begin{equation}
    U((m+1)\Delta t)|\psi \rangle=[X,U(\Delta t)]U(m\Delta t)]|\psi \rangle,
\end{equation}
and
\begin{eqnarray}
 [X,U((m+1)\Delta t)]|\psi \rangle&=&
[X,U(\Delta t)]U(m\Delta t)|\psi \rangle\nonumber\\
&&+U(\Delta t)[X,U(m\Delta t)]|\psi \rangle .
\end{eqnarray}

\begin{figure}[t]
\begin{center}\leavevmode
    \includegraphics[width=84mm]{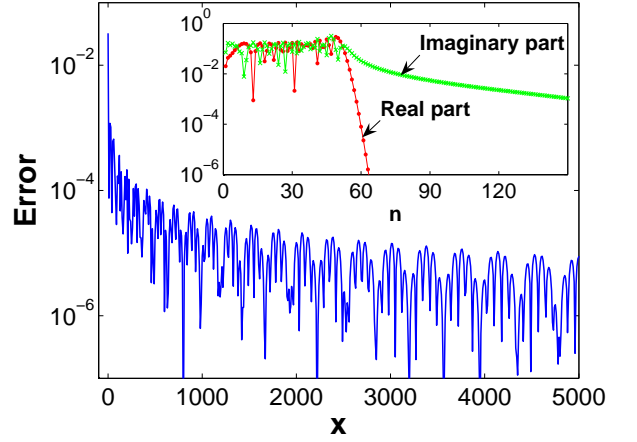}
    \caption{(Color online) Error in the modulus of the Chebyshev approximation for $f(x)=\ee^{-i\sqrt{x}}$ with $x\in[0,10^4]$, $a=5000$, $b=2500$ and $N_\mathrm{poly}=200$. In the inset, the absolute value of the real parts and the imaginary parts of the corresponding Chebyshev coefficients are shown, respectively.}
\label{Error}
\end{center}
\end{figure}
Thus, the evolution of $[X,U(t)]|\psi\rangle$ can be calculated step by step from any
starting $|\psi\rangle$ provided $c_n$ in Eq.(\ref{chebyshev coefficents}) are known. 
Accordingly, the algorithm scales linearly with the system size and the computation time. 

For $U(t)=\ee^{-i\sqrt{D}t}$, these coefficients cannot be calculated
explicitly. We therefore have to introduce a discrete grid and employ a numerical quadrature
formula. Here, we use the Chebyshev-Gauss grid of which the interpolating points are
the $\it{N}$ zeros of $Q_N(x')$:
\begin{equation}
x'_k=\cos\left(\frac{\pi\left(k+\frac{1}{2}\right)}{N}\right), k=0,1,...N-1
\end{equation}
and the related quadrature formula is 
\begin{equation}
 \int_{-1}^{1}dx'\frac{f(x')}{\sqrt{1-x'^2}}=\frac{\pi}{N}\sum_{k=0}^{N-1}f(x'_k).
\end{equation}
The calculated expansion coefficients are plotted in the inset of Fig.~\ref{Error} for $f(x)=\ee^{-i\sqrt{x}}$ with $x\in[0,10^4]$, $a=5000$ and $b=2500$.
After some steps, $c_{n}$ is seen to decay towards zero. The decay rate of the imaginary part is much slower than the real part, due to the fact the imaginary part of $\ee^{-i\sqrt{x}}$ is not differentiable at zero. 
The error of the approximated modulus with $N_\mathrm{poly}=200$ is shown in Fig.~\ref{Error}. 
The approximation keeps a good accuracy when $x$ is not close or equal to zero.
However, the error around $x=0$ is very large, which indicates the time evolution based on expansion of $U(t)$ is unstable. 
If we transform $U(t)\to\mathcal{U}(\tau)=\mathrm{e}^{-iD\tau}$, the Chebyshev expansion coefficients in Eq.~\ref{chebyshev
series} can be evaluated analytically,
\begin{eqnarray} 
c_{n}(\Delta \tau)&=&\frac{2}{\pi(1+\delta_{n,0})}\int_{-1}^{1}dx'\frac{1}{\sqrt{1-x'^2}}\ee^{-i\Delta \tau(a+2bx')}Q_{n}(x')\nonumber\\
     &=&\frac{2}{1+\delta_{n,0}}\ee^{-ia\Delta \tau}(-i)^nJ_n(2b\Delta \tau),
\end{eqnarray}
of which both the real part and the imaginary parts decay rapidly with $n$, resulting in a high accuracy of the truncation approximation for the whole spectrum.  

\section{Results and discussion}\label{section:results}
\subsection{Isotope-disordered CNT}
\begin{figure}[t]
    \begin{center}\leavevmode
    \includegraphics[width=84mm]{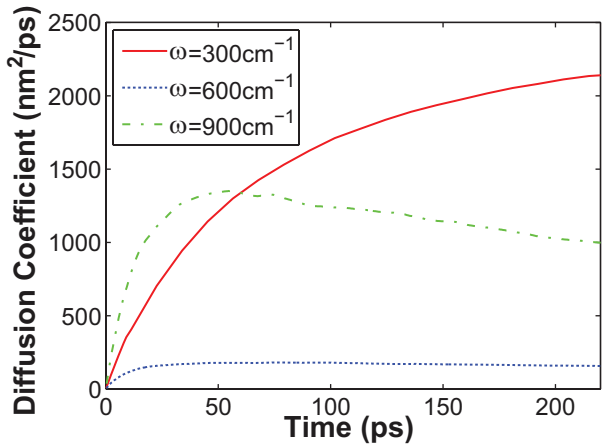}
    \caption{Time dependent diffusion coefficients $\diff(\omega,t)$ for the system of CNT with 10.7\% $^{14}$C isotope disorder at three chosen frequencies.}
    \label{2}
    \end{center}
\end{figure}
An interesting source of phonon scattering is the isotope disorder. 
In carbon-based materials (nanotubes, graphene), the controlled incorporation of $^{13}$C impurities has been experimentally demonstrated.\cite{c13,c13graphene}
The question about a possible coherent localization phenomenon in such disordered nano-structures has been discussed in Ref.~\onlinecite{Savic} using the Green's function method. 
As a test case, we investigate a CNT(7,0) with 10.7\% $^{14}$C impurities to validate our numerics in comparison with the Green's function results\cite{Savic} and also with the analytical formula for the elastic MFP,
\begin{equation}\label{AnalyticalMFP}
    \ell_\mathrm{e}(\omega)=\frac{12aN_\mathrm{uc}N_\mathrm{ch}(\omega)}{\pi^2 f|\frac{\Delta M}{\overline{M}}|^2\rho_\mathrm{uc}^2(\omega)\omega^2},
\end{equation}
where $a$ is the length of the lattice vector in the translational direction, $N_\mathrm{uc}$ is the number of atoms in each unit cell,  $\rho_\mathrm{uc}$ is the density of states per unit cell, $f$ is the percentage of 
isotopic impurities having mass difference $\Delta M$, and $\overline{M}$ is the average mass of the atoms (see Ref.~\onlinecite{Klemens,srivastava}). 
Fig.~\ref{2} displays the evolution of the wave-packet dynamics for phonon modes with different frequencies.
The linear increase of $\diff(\omega,t)$ at $t>0$ observed in all cases 
indicates ballistic transport at relatively short distances,
 whereas the decrease of $\diff(\omega,t)$ for the mode with frequency $\omega=900{\rm cm}^{-1}$ 
 at $t\geq 50$~ps is a signature of localization phenomena. 
\begin{figure}[t]
    \begin{center}\leavevmode
    \includegraphics[width=84mm]{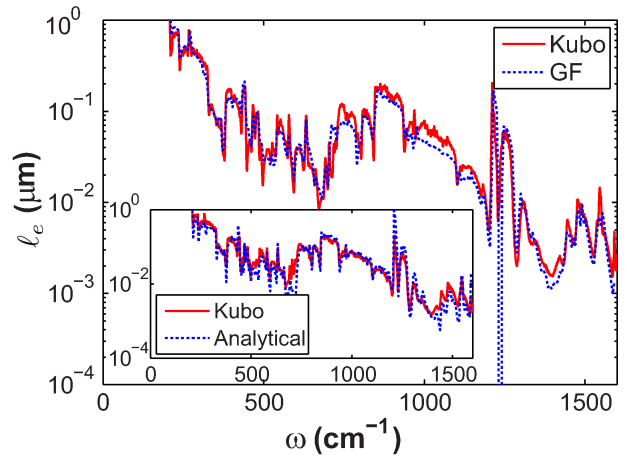}
    \caption{(Color online) Frequency-dependent elastic MFP ($\ell_\mathrm{e}$) in CNT with isotopic disorder (10\%) obtained by using the Kubo method, the GF method (data taken from Ref. \onlinecite{Savic} by courtesy), and the analytical formula.}
    \label{MFP_CNT}
    \end{center}
\end{figure}
The saturation of $\diff(\omega,t)$ to a maximum value characterizes diffusive transport.
From the saturation values, the evolution of the elastic MFP ($\ell_\mathrm{e}\simeq2\ell$) as a function of the phonon frequency and disorder features can be extracted.
The results compare well with those obtained with Green functions~\cite{Savic} (Fig.~\ref{MFP_CNT}-main frame), as well as with Eq.~(\ref{AnalyticalMFP}) (Fig.~\ref{MFP_CNT}-inset). Only at the singularities of the phonon spectrum, where disorder induces a broadening of states which impact on the numerical MFP, they differ more from those obtained with the analytical expression.

\subsection{Edge disordered graphene nanoribbons}

Graphene nanoribbons (GNRs) are strips of graphene with widths varying from a few to several tens of nanometers, depending on their fabrication processes.\cite{Graphene}
In contrast to the two-dimensional graphene, which is a zero-gap semiconductor, the narrow lateral size of GNRs entails quantum confinement effects and allows a modulation of the corresponding electronic band gap. 
The vibrational band structures are also affected by the confinement.\cite{GRAPHVIB}
Two types of GNRs with highly symmetric edge shapes, i.e. zigzag (ZGNR) and armchair (AGNR),  have been predicted and experimentally observed (see Ref.~\onlinecite{Graphene} for a review). 
Here, we consider ZGNRs of different widths with edge disorder and evaluate the corresponding transport MFPs. 
The comparison of the transport MFPs and the thermal conductances of ZGNR with AGNR having the same width is also studied.

\begin{figure}[t]
    \begin{center}\leavevmode
    \includegraphics[width=84mm]{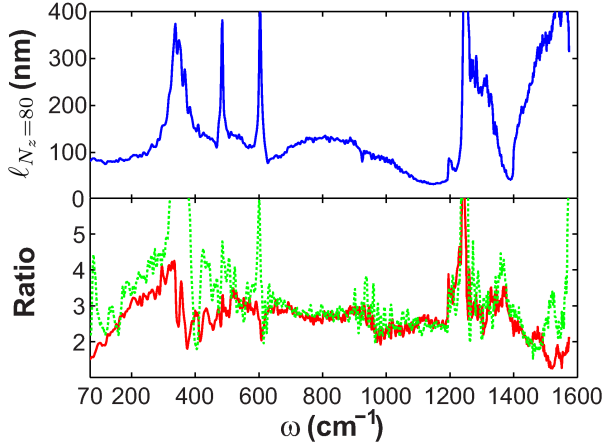}
    \caption{(Color online) Top panel: Transport MFP for ZGNR with $N_z=80$ (width of 17.04 nm) with disorder density of $10\%$. Bottom panel: Frequency-dependent MFP ratio $\ell_{N_z=80}/\ell_{N_z=40}$ (solid line) and $\ell_{N_z=40}/\ell_{N_z=20}$ (dashed line).}
    \label{MFP_ZGNR}
    \end{center}
\end{figure}

We use the fourth nearest neighbor force constants for building the dynamical matrices. \cite{saito,zimmermann}
The ribbon widths are defined with the number of zigzag chains $N_z=20$, 40, and 80 for ZGNR, and the number of dimers $N_a=138$ for AGNR.
The relative amount of edge defects (removed carbon atoms at the edges) is chosen to be  $10\%$.

Fig.~\ref{MFP_ZGNR}(top panel) shows the frequency-dependent behavior $\ell(\omega)$ of the zigzag ribbon with $N_z=80$ for $10\%$ edge disorder. 
Large modulations of $\ell(\omega)$ driven by the underlying vibrational band structure are observed. 
For a fixed disorder strength, the MFP is found to increase almost linearly with the ribbon width.
In Fig.~\ref{MFP_ZGNR} (bottom panel), the ratio $\ell_{N_{z_1}}/\ell_{N_{z_2}}$ are plotted for $N_z=20$, 40 and 80, keeping the width ratio the same.
One observes that the scaling $\ell_{N_z=80}/\ell_{N_z=40}\simeq \ell_{N_z=40}/\ell_{N_z=20}$ generally holds.
This behavior is due the fact that the scattering rate decreases with increasing width, a behavior previously reported for electron transport in both disordered CNTs and GNRs.\cite{TNT}
Since the minimum accessible frequency within a reasonable computation time is limited, $\diff_\mathrm{max}$ could not be reached for $\omega<70$~cm$^{-1}$.

\begin{figure}[b]
    \begin{center}\leavevmode
    \includegraphics[width=84mm]{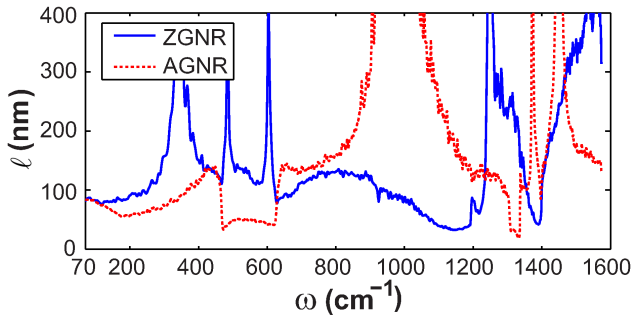}
    \caption{(Color online) Transport MFPs for the AGNR (with $N_a=138$) and the ZGNR (with $N_z=80$) at disorder density of 10\%.}
    \label{MFP_AGNR}
    \end{center}
\end{figure}

Fig.~\ref{MFP_AGNR} shows the comparison of transport MFPs in AGNR and ZGNR of approximately equal widths (17~nm) with a $10\%$ edge disorder. 
The considerable differences in the MFPs lead to an edge shape dependent thermal conductance behavior.
Different from the ZGNR, AGNR possesses a wide range of quasi-ballistic modes with MFPs as large as several $\mu$m around 950~cm$^{-1}$. 
We note that, the ballistic and diffusive modes with MFPs of several hundreds of nanometers predominate the conductance, jeopardizing any possibility to observe the onset of Anderson localization, as previously discussed for small diameter disordered carbon nanotubes \cite{Savic}, and quantum interference effects can be neglected for samples shorter than several $\mu$m.
We obtain the transmission according to $\trans(\omega)=N_\mathrm{ch}/(1+L/2\ell(\omega))$ instead of Eq.~(\ref{Trans-diffusive}), 
in order to take the contact resistance into account.
$\mathcal{T}(\omega)$ for $\omega<70$ {cm}$^{-1}$ is obtained by linearly interpolating between $\trans$ between $\omega=0$ to 70~cm$^{-1}$, where $\trans=4$ at $\omega=0$.
It has been shown that the error caused by this interpolation is less than 1.5\% for thermal conductance at room temperature.\cite{Wu1}

\begin{figure}[t]
\begin{center}\leavevmode
	\includegraphics[width=84mm]{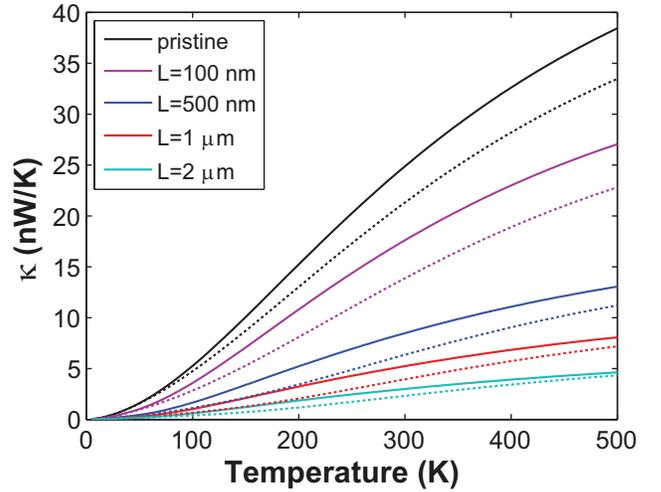}
	\caption{(Color online). Thermal conductance for the ZGNR($N_z=80$) (solid lines) and the AGNR($N_a=138$) (dashed lines) with edge disorder of $10\%$ and various ribbon lengths.}
\label{gnr_transmission}
\end{center}
\end{figure}

Thermal conductances for both systems are shown in Fig.~\ref{gnr_transmission}. 
The difference between the pristine thermal conductances of AGNR (dashed lines) and ZGNR (solid lines) is a result of the anisotropy in the phonon dispersion. 
Meanwhile, the phonon MFPs of AGNR are smaller than those of ZGNRs for the frequencies dominating the thermal conductance at low temperatures (see Fig.~\ref{MFP_AGNR}). 
These two factors cause the thermal conductance of AGNR to be smaller than that of the ZGNR for a fixed edge disorder strength and ribbon length.
Although the difference is found to be reduced as the ribbon width increases (not shown), coherent phonon propagation is still sensitive to the ribbon edge shape.

\subsection{Limits of the methodology}\label{section:limits}
The phonon wave-packet dynamics is originally determined by the time evolution operator $U(t)$, however the Chebyshev expansion of $U(t)$ has a relatively large error for low frequencies. 
Therefore we employ $\mathcal{U}(\tau)$ in the numerical calculation.
The drawback of this approximation is that, low frequency modes evolve much slower than high frequency modes, as a result, MFPs for those modes cannot be obtained within a reasonable computation time. 
Also, the accuracy of this approximation needs to be checked when the disorder in the system is relatively strong. 
In the case of very strong disorder, the approximation is less accurate, and one has to use the expansion of $U(t)$ directly. 
Even though the time evolution based on the expansion of $U(t)$ cannot give the correct information about very low frequency modes, features for the high frequency modes can still be extracted, provided that the time step and the number of Chebyshev polynomials are chosen properly.
The lowest accessible frequency by using $U(t)$ can be smaller than the one from $\mathcal{U}(\tau)$ depending on the system.

\section{Conclusion}

We have presented an efficient linear scaling approach to compute the coherent phonon wave-packet propagation in real-space and to evaluate the related thermal conductance. 
The computational accuracy and efficiency were demonstrated for isotope disordered carbon nanotubes and large width graphene nanoribbons with edge disorder, respectively.  
A strong impact of edge disorder profile on the thermal conductance was found, as well as an edge shape dependence of thermal conductance,  opening interesting perspectives for thermoelectrical applications.
One should remark that this linear scaling method can be implemented without major difficulty to a wide range of other materials, including Boron-nitride-based materials \cite{savic2} or silicon-based materials (nanowires, superlattices, etc.).\cite{mads}

\begin{acknowledgments}
This work was supported by the priority program {\it Nanostructured Thermoelectrics} (SPP-1386) 
of the German Research Foundation (DFG contract CU 44/11-1), 
the cluster of excellence of the Free State of Saxony {\it ECEMP - European Center for Emerging Materials and Processes Dresden} (Project A2),
the European Social Funds (ESF) in Saxony (research group InnovaSens), 
and the Alexander von Humboldt Foundation.
This work is also supported by the NANOSIM-GRAPHENE Project No. ANR-09-NANO-016-01 funded by the French National Agency (ANR) in the frame of its 2009 programme in Nanosciences, Nanotechnologies $\&$ Nanosystems (P3N2009) 
and by the WCU (World Class University) program sponsored by the South Korean Ministry of Education, Science, and Technology Program, Project no. R31-2008-000-10100-0.
The authors are thankful to N. Mingo for fruitful discussions. 
W.~L. thanks CAS-MPG joint doctoral promotion program. The Center for Information Services and High Performance Computing (ZIH) at the TU-Dresden is acknowledged.
\end{acknowledgments}

\end{document}